\documentstyle[12pt,epsfig]{article}                        
\newcommand{\ber}{\begin{eqnarray}}
\newcommand{\eer}{\end{eqnarray}}
\newcommand{\bea}{\begin{equation}}
\newcommand{\eea}{\end{equation}}
\newcommand{\del}{\partial}
\begin{document}
\title{\bf Thermodynamic properties of a trapped Bose gas  : A diffusion Monte Carlo study} 
\author {\bf S. Datta \\
 S. N. Bose National Centre for Basic Sciences}
\maketitle
\begin{abstract}
We investigate the thermodynamic properties of a trapped Bose gas of Rb atoms
interacting through a repulsive potential at low but finite
temperature ( $k_B T < \mu < T_c$ ) by Quantum Monte Carlo method based 
upon the generalization of Feynman-Kac method[1] applicable to many body 
systems at T=0  to finite temperatures. In this letter we report temperature 
variation of condensation fraction, chemical potential, density profile, total 
energy of the system, release energy, frequency shifts and  moment of inertia 
within the realistic potential model( Morse type ) for the first time by 
diffusion Monte Carlo technique.  The most remarkable success was
in achieving the same trend in the temperature variation of frequency shifts as
was observed in JILA[2] for both $m=2$  and $m=0 $ modes. For other things 
we agree with the work of Giorgini et al[3], Pitaevskii et al[4] and Krauth[5].
\end{abstract}
\newpage
\section{Introduction}
After the experimental realization[6] of Bose Einstein Condensation and 
measurement[7] of thermodynamic quantities in alkali gases,
there has been a growing interest in the theoretical study of these system 
since these could easily be modelled as systems with weakly interacting 
condensate and along with other things  theoretical predictions about 
thermodynamical properties were possible in terms of simple quantum 
statistical mechanics. Most of these studies involve computational 
techniques to solve the relevant many body system and based on mean field 
theory such as the Gross Pitaevskii[8] technique. Despite their success in 
explaining the ground state properties, predictions in finite temperature 
properties become only approximate and often can lead to incorrect 
predictions as we have seen from the discripencies[9-11] between theory and 
experiment in explaining JILA top data . As a matter of fact mean field 
theory breaks down near $T_c$. It is therefore necessary to develop 
alternative computational methods which can solve these many body problems 
more accurately and rigorously. 

Thermodynamics of Bose gases  was studied  before at a higher temperature 
( $k_B T >> \hbar \omega $ ) by a semiclassical treatment[3].
Since effects of interactions become more pronounced at low temperatures we 
restrict our discussions at low but finite temperature ( $k_B T < \mu < T_c$ ). At low temperature the de Broglie wavelength of the atoms become appreciable, 
the study of thermodynamic behaviour at low temperatures ( of the order of 
harmonic oscillator temperature ) requires a quantum description of a lowlying 
elementary modes. As Quantum Monte Carlo technque and many body theory are
closely connected, in this letter we present a quantum monte Carlo method 
namely Generalized Feynman-Kac method (GFK)[12,13] to study the 
thermodynamic properties of a Bose gas. From the equivalence of the
imaginary time propagator and temperature dependent density matrix, 
finite temperature results can be obtained from the same zero temperature 
code by running it for finite time. The first Monte carlo calculations [5] on 
BEC deals with temperature dependence of condensation fraction and the other 
remarkable Monte Carlo calculation[14] deals with the ground state properties. 
We calculate temperature variation of condensation fraction, 
total energy, release energy, frequency shift, chemical potential and moment of 
inertia for system of 100 $ Rb^{87} $ atoms. 
\section {Theory}
All the  thermodynamic quantities of interest are connected with the evaluation of eigensolution and eigenenergy  of the the many body system. 
In this non mean field approach,  we consider the full Hamiltonian for 
100 $Rb^{87}$ atoms interacting through Morse potential at low but finite 
temperature and the solution of corresponding Schroedinger equation in path 
integral representation. We first describe the T=0 version of
the GFK formalism[12,13] and then  generalize it to  finite temperatures.
For the Hamiltonian $H=-\Delta/2+V(x)$ consider the initial value problem
\ber
i\frac{\del u}{\del t}& =& (-\frac{\Delta}{2}+V)u(x,t)\nonumber\\
& &u(0,x)=f(x)
\eer
with $x \in  R^d$ and $u(0,x)=1$. The solution of the above equation can
be written in Feynman-Kac representation as
\bea
u(t,x)=E_xexp\{-\int_0^t V(X(s))ds\}
\eea
where X(t) is a Brownian motion trajectory and E is the average value of the
exponential term with respect to these trajectories. To  speed up the 
convergence one can incorporate importance sampling in the underlying 
stochastic process and the lowest energy formula for
eigenvalue for a  given symmetry obtained from the large deviation
priniciple of Donsker and Varadhan [15] can be written as 
\bea
\lambda={\lambda}_T-\lim_{t\rightarrow \infty}
 \frac{1}{t}ln E_xexp\{-\int_0^t V_p(Y(s))ds\}
\eea
where Y(t) is the diffusion process which solves the stochastic differential
equation
\bea
dY(t)=\frac{{\Delta}{\psi}_T(Y(t))}{{\psi}_T(Y(t))}dt+dX(t)
\eea
                                                                           
The temperature dependence comes from the realization that
the imaginary time propapagator is identical to the temperature
dependent density matrix  if $t\Rightarrow\beta=  1/T$
holds.
        
This becomes obvious when we consider the eqs[16]
\bea
-\frac{\del k(2,1)}{\del t_2}= H_2k(2,1)
\eea
and
\bea
-\frac{\del \rho}{\del \beta}=H_2 \rho(2,1)
\eea
and compare
\bea
k(2,1)=\sum_i\phi_i(x_2){\phi_i}^*(x_1)e^{-(t_2-t_1)E_i}
\eea
and
\bea
\rho(2,1)=\sum_i\phi_i(x_2){\phi_i}^*(x_1)e^{-\beta E_i}
\eea
For Zero temp FK we had to extrapolate to $t\Rightarrow \infty$.
For finite run time t in the simulation, we have finite temperature results.
Here we show how to change our formalism to go from zero to finite
temperature. We begin with the definition of finite temperature.
A particular temperature 'T' is said to be finite if
$\Delta E < kT$ holds.
The temperature dependent density matrix can be written in the following form
\bea
\rho(x,x^{\prime},\beta)={\rho}^{(0)}(x,x^{\prime},\beta)\nonumber\\
\times < exp[-\int_{0}^{\beta}V_p[X(s)]ds]>_{DRW}
\eea
At finite temperature thus free energy can be written as
\bea
F=
-ln Z(x,\beta)/\beta=-{ln Z^{0}(x,\beta)}/\beta-{ln  < exp[-\int_{0}^{\beta}
V_p[X(s)]ds]>_{DRW}}/\beta
\eea
\newpage
\subsection{The stationary state of the condensate} 

To calculate the condensate energy and condensate density, we consider a cloud
of N atoms interacting through repulsive potential and placed in a three
dimensional anharmonic oscillator potential. We will assume that the condensate
system has relaxed to a stationary state and at low energy the time independent
Schroedinger equation representing the stationary state can be written as  
\bea
[-{\Delta}/2+V_{int}+\frac{1}{2} \sum_{i=1}^N [{x_i}^2
 +  {y_i}^2+\lambda{z_i}^2]]{\psi}_0(\vec{r})
=E{\psi}_0(\vec{r})
\eea
where $\frac{1}{2} \sum_{i=1}^N [{x_i}^2
 +  {y_i}^2+\lambda{z_i}^2] $ is the anisotropic
potential with anisotropy factor $\lambda=\frac{{\omega}_z}{{\omega}_x}$.
Now
 \bea
 V_{int}=V_{Morse}=
\sum_{i,j} V(r_{ij})=\sum_{i<j}D[e^{-\alpha(r-r_0)}(e^{-\alpha(r-r_0)}-2)]
 \eea
Here we assume that the condensate oscillates in a static thermal bath.
There is no interaction between the condensate and the thermal bath.
The principal effect of finite temperature on the excitations is the depletion
of condensate atoms. In the dilute limit and at very low energy only binary 
collisions are possible and no three body recombination is allowed. 
In such two body scattering at low energy first order Born approximation is 
applicable and the interaction strength 'D' in the dimnsionless form($\gamma$) 
turns out 
to be 
\bea
\gamma=4\frac{a{\alpha}^3}{se^{\alpha r_0}(e^{\alpha r_0}-16)}
=4.9 \times 10^{-5}
\eea
For more details, one should look at [17].
In the above expression 
$ a $ is the scattering length of Rb, $ \alpha $ is the width of the Morse
potential, $r_0$ is the minimum of the potential well, 's' is the length scale
in units of harmonic oscillator and $\lambda$ is the anisotropy factor.
 Here we have used[17]                                             
$\alpha=.29$ in harmonic oscillator units, $r_0=9.67$ in harmonic oscillator units, $ a=52 \times 10^{-10}$ m, 
$s=.12\times 10^{-5}$ m and $\lambda=\sqrt 8$
For more details, one should look at [17].
\newpage
\subsection {\bf Effect of noncondensate}
In the case of noncondensate the system can be considered as a thermal gas.
To calculate noncondensate energy and density we need to study the effect of 
noncondensate explicitly and consider
the following stationary state for the thermal gas.
\bea
[-{\Delta}/2+2V_{int}+V_{trap}]\psi_j(\vec{r})
=E_{nc}\psi_j(\vec{r})
\eea
\bea
[-{\Delta}/2+2V_{int}+\frac{1}{2} \sum_{i=1}^N [{x_i}^2
 +  {y_i}^2+\lambda{z_i}^2]\psi_{j}](\vec{r})
=E_{nc}\psi_{j}(\vec{r})
\eea
The basis wavefunction ${\psi}_{j}$ which describes the noncondensate should
be chosen in such a way that it is orthogonal to ${\psi_0}$ as in Eq.(11)
The most common way to achieve a orthogonal basis in Schroedinger prescription 
is to consider the dynamics of noncondensate in an effective potential[18,19]
$V_{eff}=V_{trap}+2V_{int}$. The factor 2 represents the exchange
term between two atoms in two different states. The energy in the case of
lowest lying modes  then corresponds to $E=E_c+E_{nc}$. One can calculate 
the $E_{nc}$ using the same parameters as discussed in Sec 2.1
\newpage
\section{Results and discussions}
\subsection{\bf The effect of temperature on frequency shifts}
\subsubsection {\bf m=2 mode}
Using Eq.(3), one can calculate the lowest lying energies due to any symmetry. 
To calculate the frequencies for $m=0 $ and $m=2$ modes, one needs to 
find the energy differences of each of these states and the ground state.
Underneath we show the data for frequency shifts for both $m=2$ and $m=0$ modes.
For $m=2$ mode, considering motion of condensate only we achieve the 
downward shift of data all the way to $T=0.9T_c$. 
But for the $m=0$ mode we need to consider the dynamics of thermal cloud also 
as discussed above.
\vskip 1cm
\begin{figure}[h!]
\centering
\epsfxsize=3.5in{\epsfbox{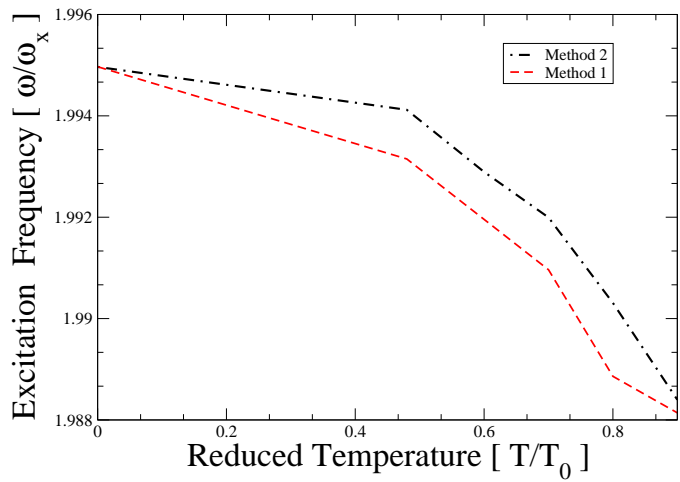}}
\caption{Effects of temperature on m=2 mode; this work.
The top cuve from equivalent T=0 system[method 2], the bottom curve by putting
temperature directly[method 1]. Both show agreement with JILA experimental 
data[2] all the way  up to $0.9T_c$  and Ref[20,21]}
\end{figure}
\newpage
\subsubsection {\bf m=0  mode}
For $m=0$ mode, considering the dynamics of condensate ( Sec 2.1) alone we do 
not get the upward shift as observed in  JILA experimental data. But when the 
motion of thermal cloud (Sec 2.2) is considered in a dynamical manner, we 
observe the expected upward shift at around $T=0.7T_c$.
\vskip 1.0cm
\begin{figure}[h!]
\centering
\epsfxsize=4in{\epsfbox{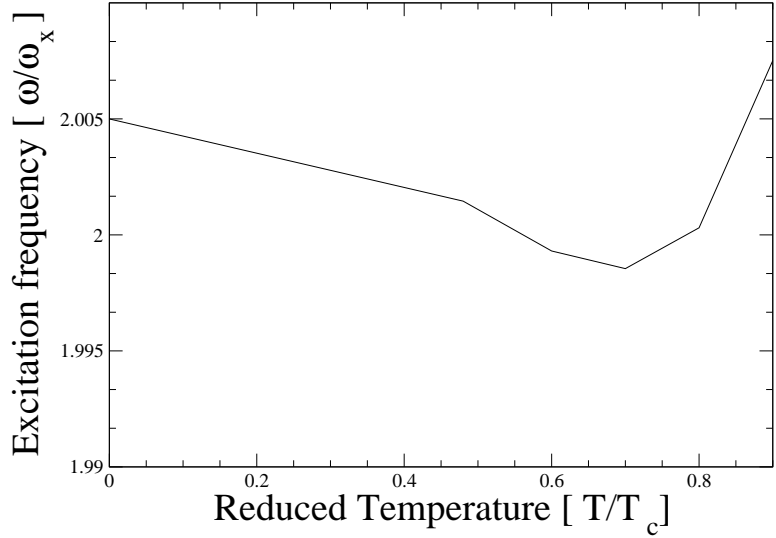}}
\caption{Effects of temperature on m=0 mode from GFK considering noncondensate
dynamics[this work], shows resemblence with JILA[2] and Ref[20,21] }
\end{figure}
\vskip 1cm
\subsection{ The effect of temperature on condensate density }
The condensate density can be evaluated solving Eq.(11) and using Eq.(9).
Underneath we plot the axial density due to condensate along x axis.
\vskip 0.5 cm
\begin{figure}[h!]
\centering
\epsfxsize=3.5in{\epsfbox{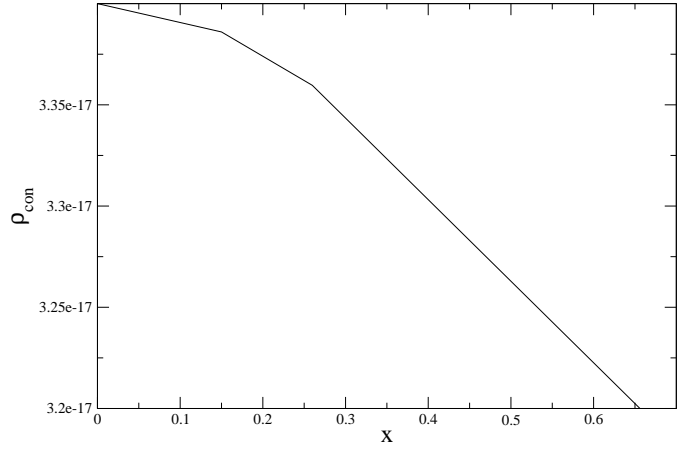}}
\caption{Axial density profile due to Condensate  at temperature T=0.48}
\vskip 2cm
\end{figure}
\begin{figure}[h!]
\centering
\epsfxsize=3.5in{\epsfbox{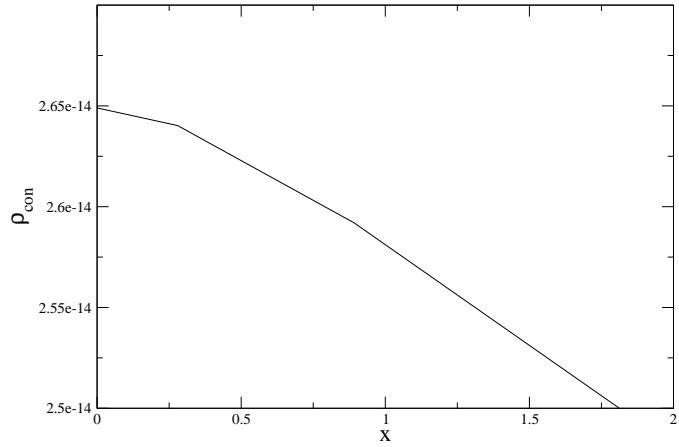}}
\caption{Axial density profile due to condensate at temperature T=0.6}
\end{figure}
\newpage
Fig 3, 4 and 5 are concerned with
condensate density profile vs x for temperatures $T=0.48$, $T=0.6$ and $T=0.7$
respectively for a system of 100 Rb atoms.
We see that the center density for condensate increases  as temperature is
increased. This agrees with the earlier work of Krauth[5]
\vskip 1cm
\begin{figure}[h!]
\centering
\epsfxsize=4.0in{\epsfbox{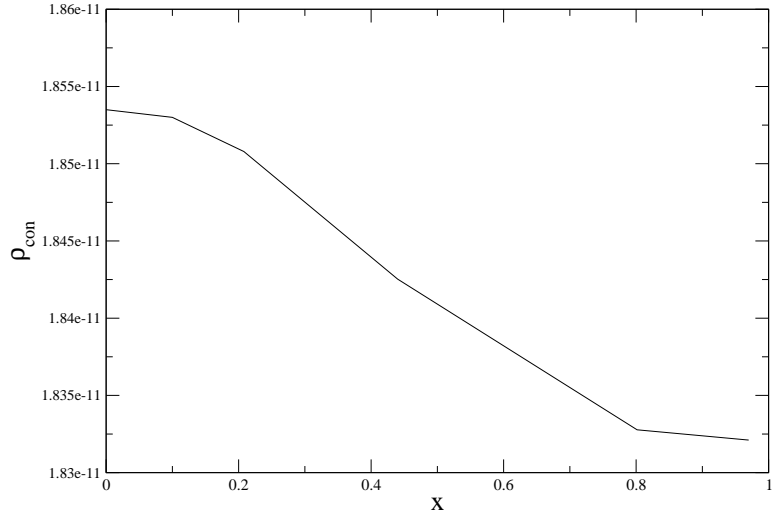}}
\caption{Axial density profile due to condensate  at temperature T=0.7}
\end{figure}
\newpage
\subsection{\bf The effect of temperature on the total density}
Following the theory in Sec 2.1 and 2.2  and using Eq.(9) one can calculate 
the total density of the Bose gas. Underneath we show the temperature variation of the total density along the x axis.
\vskip 1 cm
\begin{figure}[h!]
\centering
\epsfxsize=3.5in{\epsfbox{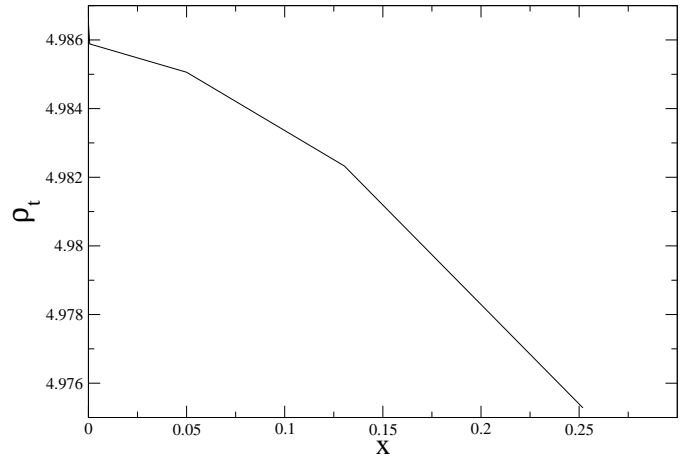}}
\caption{total density profile at temperature T=0.48}
\vskip 0.5cm
\end{figure}
\begin{figure}[h!]
\centering
\epsfxsize=3.5in{\epsfbox{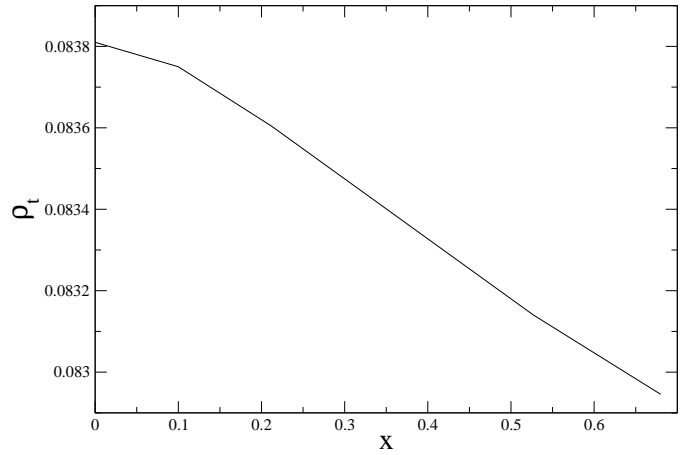}}
\caption{total density profile at temperature T=0.6}
\vskip 2cm
\end{figure}
\begin{figure}[h!]
\centering
\epsfxsize=3.5in{\epsfbox{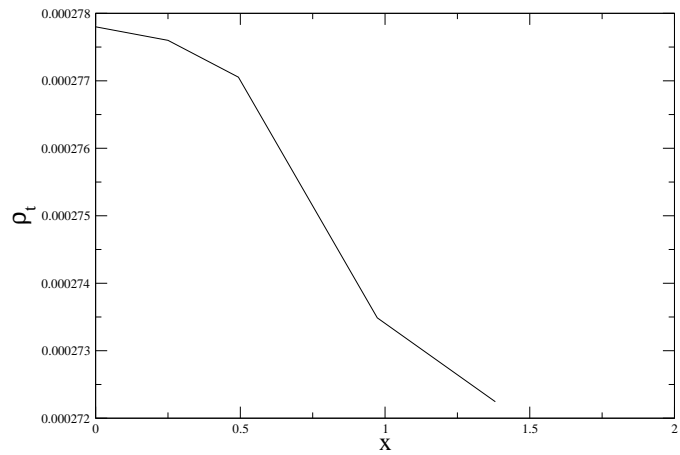}}
\caption{total density profile at temperature T=0.7}
\end{figure}
\newpage
On the other hand in Fig 6, 7 and 8 which represent the total density 
profile vs x at $T=.48$,$T=0.6$ and $T=0.7$ respectively, we see the opposite 
trend as the center density decreases as the temperature is raised.
This also is in agreement with the ealier observation by Krauth[5]
\newpage
\subsection {\bf Effect of temperature on the total energy of the Bose gas}
{\bf Total energy of the system}
The condensate and noncondensate energies can be evaluated solving Eq. (11) and
Eq.(15) and  using Eq(3). Then combining condensate and noncondensate energy 
we get the total energy of the system. The total energy of condensed and 
noncondensed component  of a trapped Bose gas is a combination of $E_{kin}$,
$E_{ho}$ and $E_{int}$ for each component separately. The trend in the 
temperature variation of total energy is found to be the same as in [3,4]. 
In principle, specific heat can be calculated as the temperature derivative of 
total energy/particle keeping the confining potential constant. We will report 
it for bigger systems somewhere else. Release energy can be represented defined to be the energy obtained after switching off trap. We get similar trend the 
temperature variation of release energy as we see in ref[4]  \\
$E_{rel}= E_{kin}+E_{int}$\\
\vskip 1.5cm
\begin{figure}[h!]
\centering
\epsfxsize=3.5in{\epsfbox{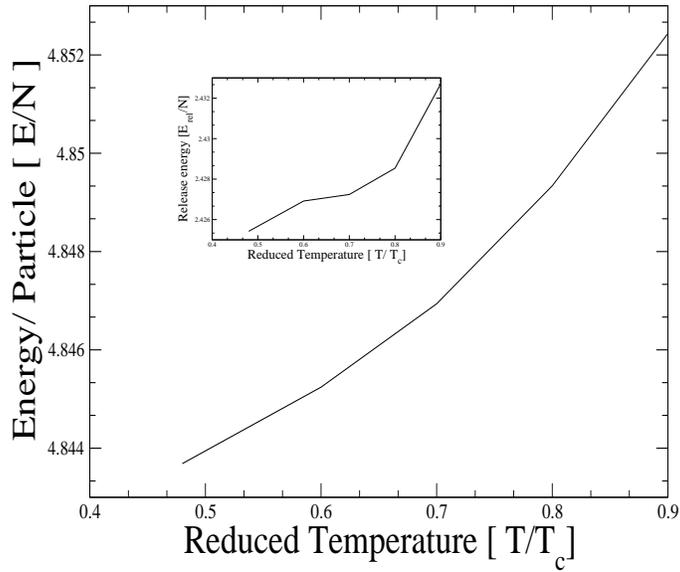}}
\caption{Total energy/particle as a function of reduced temperature.
Inset: Release energy as a function of temperature}
\end{figure}
\newpage
\subsection{\bf Effects of temperature on condensation fraction}
 Interaction lowers the condensation fraction ($\frac{N}{N_c}$ ) for repulsive 
potentials. Some particles always leave the trap because of the repulsive 
nature of the potential and on the top of it, if temperature is increased 
further, more particles will fall out of the trap and get thermally distributed.This decrease in condensation fraction eventually would cause the shifts in the 
critical temperature ( $T_c$ decreases ). We would observe this in (Fig.10 ). 
Earlier this was done by W. Krauth[5] for a large number of atoms by path 
integral Monte Carlo method.
\vskip 0.5cm
\begin{figure}[h!]
\centering
\epsfxsize=3.5in{\epsfbox{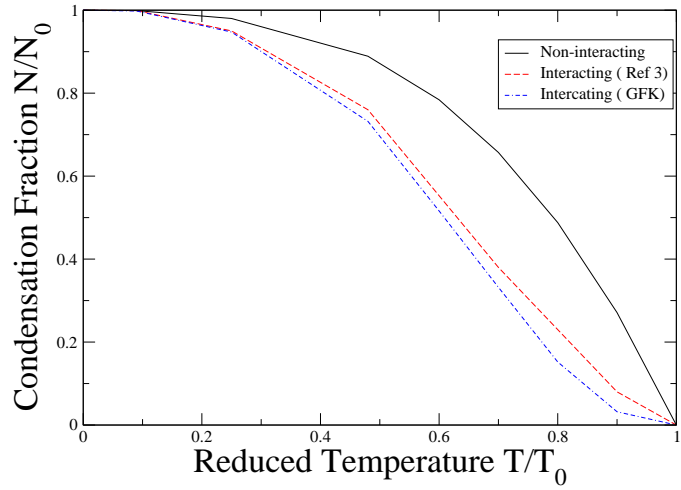}}
\caption{Condensation fraction vs Reduced Temperature
; this work. The innermost curve corresponds to the 100 interacting atoms and 
the outermost curve corresponds to the noninteracting case. The middle curve 
corresponds to the interactive case in ref[3]. 
The number of condensed particles decreases with the interaction }
\end{figure}
\newpage
\subsection{\bf Effects of temperature on Chemical potential}
The chemical potential $\mu$ can be written in terms of different contributions
to the energy, namely $E_{kin}$, $E_{ho}$ and $E_{int}$ as follows[4].
\bea
\mu=\frac{1}{N}(E_{kin}+E_{ho}+2E_{int})
\eea
\begin{figure}[h!]
\centering
\epsfxsize=3.5in{\epsfbox{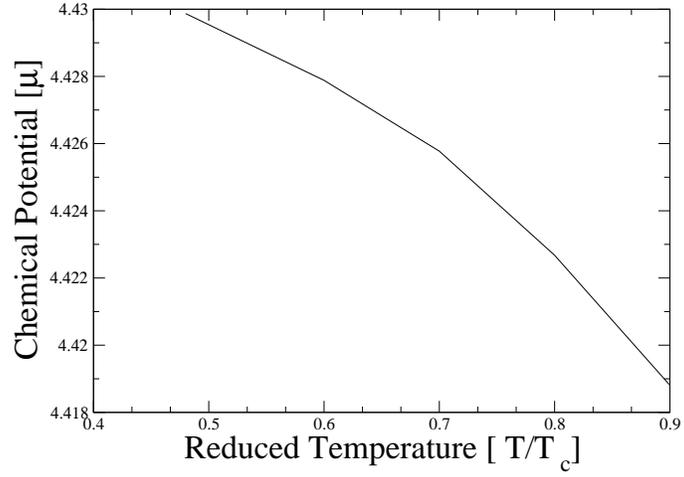}}
\caption{Chemical Potential vs Reduced Temperature}
\end{figure}
\newpage
\subsection{\bf Effects of temperature on moment of inertia }
{\bf Moment of inertia }
The deviation of moment of inertia from its rigid value is given by the
useful expression[22]
\bea
\frac{\theta}{\theta_{rig}}=\frac{N_T<r^2>_T}{N_0<r^2>_0+N_T<r^2>_T}
\eea
where $<>_0$ and $<>_T$ denote the average taken over the condensate and 
noncondensate densities of the Bose gas respectively.
\vskip 1cm
\begin{figure}[h!]
\centering
\epsfxsize=3.5in{\epsfbox{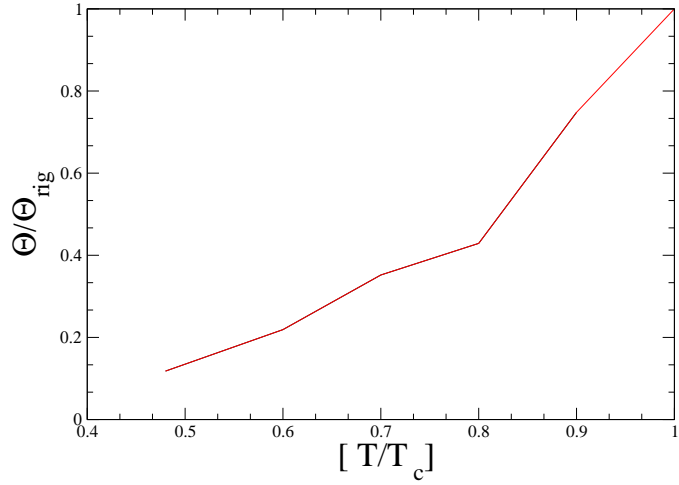}}
\caption{Moment of inertia $\theta$ divided by its rigid value $\theta_{rig}$
as a function of $T/T_c$}
\end{figure}
At $T=0$, $N_T=0$ and $\frac{\Theta}{\Theta_{rig}}=0$. On the other hand 
fo $T=T_c$, $\frac{\Theta}{\Theta_{rig}}=1$ as $N_0=0$. 
\newpage
\section{Conclusions:}
For the first time we have calculated finite temperature properties beyond 
meanfield approximation( GP etc) by Quantum Monte Carlo technique. We have 
calculated spectrum of Rb gas by considering  realistic potentials like 
Morse potential etc. instead of conventional pseudopotentials for the first 
time.  Since a dilute gas consisting of $Rb^{87}$ atoms is a 
bosonic system the random walk in GFK is exact in the limit scale, time for 
walk and the number of walks get arbitraly large and it turns out that the
this  method is a potentially good candidate for  a sampling procedure for Bose 
gases at all temperatures. 

We are dealing with only 100 atoms. Nonetheless we have been able to 
show the variation of condensate and noncondensate density with 
temperature as a hallmark of BEC and lowering of condensation fraction in 
the case of interacting case which is very unique with a system of trapped 
atoms compared to uniform system.

In our non mean field study for finite temperature excitaion spectrum for m=2 
mode, we see agreement with experimental study all the way to $ T=0.9T_c $ 
[Fig. 1]. For $ m=0 $ mode, considering the motion of thermal cloud explicitly, we observe the upward shift of data[Fig 2] in JILA[2] and Ref[20,21]. 
Since at low temperature we solve the many body problem by considering the full Hamiltonian with realistic potential full quantum mechanically and 
nonperturbatively, obviously the above modes are collective in nature and 
correspond to the m=2 and m=0 experimental modes[7].  

Since we are dealing with very small number of particles we cannot compare our 
data with other existing results directly on the same graph. We can neither use scaling method since we are working on a small system at a low temperature 
regime $k_B<\mu< T_c$. We can study and compare
the temperature variation of different quantities such as  total energy, 
release energy, moment of inertia with those in the literature[3,4,5]
only qualitatively.
 
The method is extremely easy to implement and our fortran code at this point 
consists of about 270 lines. We employ an algorithm which is essentially 
parallel in nature so that eventually we can parallelize our code and 
calculate thermodynamic properties of bigger systems with the scaling property
( of the order of 2000 atoms ) taking advantage of the new computer 
architechtures This work is in progress. We are continuing on this problem and 
hope that this technique will inspire others to do similar calculations. 

\newpage
{\bf Acknowledgements}:\\
Financial help from DST ( under Young Scientist Scheme
(award no. SR/FTP/-76/2001 )) is gratefully acknowledged. The author would like
to thank Prof J. K. Bhattacharjee, Indian Association for the Cultivation
of Science, India for suggesting the problem and many stimulating discussions.
\end{document}